\documentstyle[12pt]{article}
\begin{document}
\newcommand{\half}{\frac{1}{2}}
\newcommand{\ehalf}{\frac{e}{2}}
\newcommand{\ihalf}{\frac{i}{2}}
\newcommand{\khalf}{\frac{\kappa}{2}}
\newcommand{\iehalf}{\frac{ie}{2}}
\newcommand{\ghalf}{\frac{g}{2}}
\newcommand{\quart}{\frac{1}{4}}
\newcommand{\kquart}{\frac{\kappa}{4}}
\newcommand{\G}{\Gamma}
\renewcommand{\O}{\Omega}
\renewcommand{\S}{\Sigma}
\newcommand{\U}{\Upsilon}
\renewcommand{\a}{\alpha}
\renewcommand{\b}{\beta}
\newcommand{\e}{\epsilon}
\newcommand{\g}{\gamma}
\newcommand{\k}{\kappa}
\renewcommand{\l}{\lambda}
\renewcommand{\t}{\theta}
\newcommand{\pdp}{\Phi^{\dag}\Phi}
\newcommand{\opar}{\bar\eta}
\newcommand{\ze}{\bar{z}}
\newcommand{\cA}{\cal A}
\newcommand{\cB}{\cal B}
\newcommand{\cH}{\cal H}
\newcommand{\dmu}{\partial_{\mu}}
\newcommand{\dmup}{\partial^{\mu}}
\newcommand{\dphi}{\partial\phi}
\newcommand{\dchi}{\partial\chi}
\newcommand{\aslash}{\not\!\!\cA}
\newcommand{\hslash}{\not\!\!\cH}
\newcommand{\Dslash}{\not\!\! D}
\newcommand{\dslash}{\not\!\partial}
\newcommand{\pau}{\vec{\tau}}
\newcommand{\tsq}{\theta^2}
\newcommand{\otsq}{\bar\theta^2}
\newcommand{\ts}{\bar\theta\theta}
\newcommand{\tu}{\t\sigma^{\mu}\tt}
\renewcommand{\tt}{\bar\theta}

\begin{titlepage}
\rightline{Preprint DF/UFPB 11/97}
\rightline{La Plata-Th 97/11}
\rightline{hep-th/9707016}
\vskip 13mm
\centerline{\Large \bf Kinks Inside Supersymmetric Domain Ribbons}
\vskip 9mm
\centerline{J. D. Edelstein and M. L. Trobo}
\vskip 1mm
\centerline{{\it Departamento de F\'{\i}sica, Universidad Nacional de La Plata}}
\centerline{{\it CC 67, 1900 La Plata, Argentina}}
\vskip 5mm
\centerline{F. A. Brito and D. Bazeia}
\centerline{{\it Departamento de F\'{\i}sica, Universidade Federal da Para\'{\i}ba}}
\centerline{{\it Caixa Postal 5008, 58051-970 Jo\~ao Pessoa, Para\'{\i}ba, Brazil}}
\vskip 15mm
\centerline{\sc Abstract}
\vskip 11mm

\noindent
We study a variety of supersymmetric systems describing sixth-order
interactions between two coupled real superfields in $2+1$ dimensions. 
We search for BPS domain ribbon solutions describing minimun energy 
static field configurations that break one half of the supersymmetries.
We then use the supersymmetric system to investigate the behavior of 
mesons and fermions in the background of the defects. In particular,
we show that certain BPS domain ribbons admit internal structure in the
form of bosonic kinks and fermionic condensate, for a given range of
the two parameters that completely identify the class of systems.
\vskip 2cm

\end{titlepage}
\newpage

\section{Introduction}

The idea of topological defects that present internal structure was first
introduced in \cite{wit85}, within the context of modeling superconducting
strings. It was also explored by other authors \cite{lsh85,mac88} in 
different contexts, and more recently the works \cite{mor94,mor95}
also investigate systems that admit the existence of defects inside 
topological defects. General features of the
works just mentioned are that they consider systems in $3+1$ dimensions, and
that the potential describing the scalar fields depends on several
parameters. In this case, solutions representing topological defects with
internal structure only appear after adjusting some of the several
parameters that defines the model under consideration.

~

In some recent works \cite{bds95,bsa96,brs96a}, solitons that
emerge in certain bidimensional systems of coupled real scalar fields
were studied. The class of systems considered in those works has
been shown \cite{brs96b,bmo97} to admit a natural embedding into the
bosonic sector of a supersymmetric theory, in such a way that the set
of free parameters is quite restricted. Within this context, the existence of
topological defects inside domain walls in a model of two real scalar
field belonging to the above mentioned class of systems was first
considered in \cite{brs96b}. There, the system is defined by a 
potential that contains up to the quartic power in the scalar fields, 
and the set of parameters is reduced to just two parameters. The system
has a domain wall solution that traps in its interior a topological
defect produced by the remaining scalar field, provided the parameters 
ratio is positive definite. It also has solutions known as 
domain ribbons inside a domain wall, and they resemble a stringlike
configuration that can be either infinitely long or in the form of
a closed loop.

~

As pointed out in \cite{brs96b}, the reduction in the number of free
parameters may perhaps lead to a clearer understanding of the physical
properties these kind of systems can comprise. This is one of the main
motivation of the present paper, in which we shall further explore the
possible existence of defects inside topological defects in systems
belonging to the class of systems already introduced in \cite{bds95,brs96a}.
Here, however, we shall investigate systems containing up to the sixth
power in the scalar fields. In this case we shall restrict our investigation
to 2+1 dimensional spacetime, and so we are going to search for kinks
inside supersymmetric domain ribbons. There are many reasons to consider
such systems, among them we would like to single out the following:
Potentials with sixth-order interactions define systems that admit the
existence of solitons of different nature, at least in the sense that they
may connect adjacent vacua in a richer set of vacuum states. As we are 
going to show below, there are at least three systems that seems to be
worth investigating, one of them was already considered in \cite{bds95},
in the 1+1 dimensional case, and the others will be defined below.

~

The investigations are organized as follows. In the next Section we 
perform the construction of the class of systems of our interest
in the framework of supersymmetric field theory. In Sec.~{\ref{sec:bps}},
we show that these systems present a Bogomol'nyi bound for the energy
whose saturation is achieved provided the fields solve a set of
first-order equations, simpler than the usual equations of motion. 
The solutions of these Bogomol'nyi equations break one
half of the supersymmetries and then belong to a short supermultiplet.
In Sec.~{\ref{sec:systems}} we illustrate the procedure by investigating
some specific systems. Sec.~{\ref{subsec:system1}} is devoted to the study
of a model already introduced in \cite{bds95}, and there we show that this
system does not allow the formation of kinks neither the trapping of
mesons inside the domain ribbon. This result can be traced up to an
asymmetry produced by the fact that the domain ribbon found for the
sixth order system connects the symmetric vacuum state with
non-symmetric ground states. Nevertheless, the effective potential
owed to this configuration favor the entrapping of Majorana
fermions inside the domain ribbon. In Sec.~{\ref{subsec:system2}}
we present a simple extension of the previous 
model that circumvent the above mentioned asymmetry allowing 
the existence of a kink inside the supersymmetric domain ribbon. 
We briefly discuss on the formation of these kinks and their stability.
The third model is presented in Sec.~{\ref{subsec:system3}}, and its
main feature is that both superfields present equal footing
from the beginning. Then, we show that the system admits domain ribbons
with an internal kink of the same sixth-order nature, for a given range of 
parameters, and we briefly discuss its classical stability.
Also, we investigate the behaviour of fermions in the background of
these solutions in Sec.~{\ref{subsec:fermions}}. We end this paper in
Sec.~{\ref{sec:comments}}, where we introduce some comments,
further remarks and conclusions.

\section{Supersymmetric Systems of Coupled Real Scalars}
\label{sec:trisystems}

A general system of two real scalar fields in $2+1$ spacetime is
described by the Lagrangian density
\begin{equation}
{\cal L}=
\frac{1}{2}\partial_{\mu}\phi\partial^{\mu}\phi+
\frac{1}{2}\partial_{\mu}\chi\partial^{\mu}\chi-
U(\phi,\chi),
\end{equation}
where $U(\phi,\chi)$ is the potential, in general a nonlinear function 
of the two fields $\phi$ and $\chi$ involving several coupling constants 
for the different terms. As explained above, an interesting framework
that highly restricts the dimensionality of the parameter space is
provided by supersymmetry. In that respect, let us start by considering a 
supersymmetric field theory in $(2+1)$ spacetime dimensions, entirely 
constructed from two real superfields \cite{comment} $\Phi$ and $\Xi$, 
\begin{equation}
\Phi = (\phi,\psi,D_{\phi}) \;\;\;\;\;\;\; \Xi = (\chi,\rho,D_{\chi}) ~,
\label{eq:smult}
\end{equation}
where $\psi$ and $\rho$ are Majorana two-spinors, while $D_{\phi}$ and
$D_{\chi}$ are bosonic auxiliary fields. The Lagrangian density
can be written in terms of the superfields as
\begin{equation}
{\cal L}_{N=1} = \frac{1}{2} \int d^2\theta \left[ \bar{\cal 
D}\Phi{\cal D}\Phi + \bar{\cal D}\Xi{\cal D}\Xi + {\cal
W}(\Phi,\Xi) \right] ~.
\label{eq:lagsusy1}
\end{equation}
Here we are following the conventions introduced in \cite{ggr} and, 
as usual, ${\cal D}$ is the supercovariante derivative
\begin{equation}
{\cal D} = \partial_{\tt} + i\tt\g^{\mu}\dmu ~, 
\label{eq:super}
\end{equation}
with the $\gamma$-matrices being represented by $\gamma^0 = \tau^3$, 
$\gamma^1 = i\tau^1$ and $\gamma^2 = -i\tau^2$. 
In terms of component fields, after replacing the auxiliary fields
$D_{\phi}$ and $D_{\chi}$ by their algebraic equations of motion,
the Lagrangian density can be written as
\begin{eqnarray}
{\cal L} & = & \half\dmu\phi\dmup\phi + \half\dmu\chi\dmup\chi
+ \ihalf\bar\psi\dslash\psi + \ihalf\bar\rho\dslash\rho -
\half\left(\frac{\partial{\cal W}}{\partial\phi}\right)^2 
\nonumber \\
& - & \half\left(\frac{\partial{\cal W}}{\partial\chi}\right)^2 
- \half\frac{\partial^2{\cal W}}{\partial\phi^2}\bar\psi\psi
- \half\frac{\partial^2{\cal W}}{\partial\chi^2}\bar\rho\rho -
\frac{\partial^2{\cal W}}{\partial\phi\partial\chi}\bar\psi\rho ~,
\label{eq:lagsusy2}
\end{eqnarray}
where it is explicit that the Yukawa couplings as well as the
scalar potential of the theory entirely depends on the superpotential
${\cal W}$. In fact, we stress that the scalar potential results to be
\begin{equation}
U(\phi,\chi) = \half\left(\frac{\partial{\cal 
W}}{\partial\phi}\right)^2 + \half\left(\frac{\partial{\cal 
W}}{\partial\chi}\right)^2 ~,
\label{eq:pot}
\end{equation}
in direct correspondence with a general class of systems that comprises
quite interesting properties as was already described in 
Refs.\cite{bds95,bsa96,brs96a,brs96b}. The set of transformations 
that leave invariant the system described by $(\ref{eq:lagsusy2})$ is
\begin{equation}
\delta_{\eta}\phi = \bar\eta\psi \;\;\;\;\;\;\;
\delta_{\eta}\chi = \bar\eta\rho ~,
\label{eq:traf1}
\end{equation}
\begin{equation}
\delta_{\eta}\psi = \left(-i\dslash\phi + \frac{\partial{\cal 
W}}{\partial\phi}\right)\eta ~,
\label{eq:traf2}
\end{equation}
\begin{equation}
\delta_{\eta}\rho = \left(-i\dslash\chi + \frac{\partial{\cal 
W}}{\partial\chi}\right)\eta ~,
\label{eq:traf3}
\end{equation}
where the infinitesimal parameter $\eta$ is a real 
spinor. 
Let us now focus upon the classical configurations
of this system. We will then set the fermion fields to zero and
look for purely bosonic field configurations which, from the point
of view of the supersymmetric theory, can be understood as 
background solutions. Responses of the fermion fields to these
backgrounds can then be investigated. We then introduce the following 
useful notation: Given a functional ${\cal F}$ depending both on 
bosonic and fermionic fields, we will use ${\cal F}\vert$ to refer to 
that functional evaluated in the purely bosonic background,
\begin{equation}
{\cal F}\vert \equiv {\cal F}\vert_{\rho,\psi=0} ~.
\label{eq:spb}
\end{equation}
Under condition (\ref{eq:spb}), the only non-vanishing supersymmetric 
transformations that leave invariant the Lagrangian (\ref{eq:lagsusy2})
are those corresponding to fermionic fields.

\section{BPS Domain Ribbons}
\label{sec:bps}

Let us now show that the supersymmetric nature of the system imposes 
lower bounds for the mass per unit length of a generic bosonic static 
configuration that is homogenous in one of the spatial coordinates. 
Indeed, one can compute the conserved supercharge that generates the 
transformations $(\ref{eq:traf2})$ and $(\ref{eq:traf3})$ to be
\begin{equation}
{\cal Q}_{\a} = \half\int d^2x\left\{\left[-i\dslash\phi 
+ \frac{\partial{\cal W}}{\dphi}\right]\psi_\a + 
\left[-i\dslash\chi + \frac{\partial{\cal 
W}}{\dchi}\right]\rho_\a\right\} ~,
\label{eq:supcharge}
\end{equation}
and use it to construct the supercharge algebra over the static
bosonic background resulting
\begin{equation}
\{{\cal Q}_\a,{\cal Q}_\b\}\vert = \gamma^0_{\a\b}M + 
\e_{\a\b}Z ~,
\label{eq:algebra}
\end{equation}
where $M$ is the mass of the purely bosonic configuration
\begin{equation}
M = \half \int d^2x \left[(\vec{\nabla}\phi)^2 + 
(\vec{\nabla}\chi)^2 + \half\left(\frac{\partial{\cal 
W}}{\partial\phi}\right)^2 + \half\left(\frac{\partial{\cal
W}}{\partial\chi}\right)^2 \right] ~,
\label{eq:mass}
\end{equation}
whereas the `central extension' of the algebra $Z$ is given by
a line integral over a curve $\G$ that encloses the region
where the fields carry a finite energy density
\begin{equation}
Z = \oint_\G \vec{\nabla}{\cal W}\cdot{d\vec{x}} ~.
\label{eq:central}
\end{equation}
It is clear from its expression that the `central extension'
$Z$ is in general forced to vanish as a consequence of the
scalar nature of the real superpotential. Indeed, the 
supersymmetry algebra of any three-dimensional system with
an $N=1$ invariance does not have room for the introduction
of a central charge.

~

There is a breakthrough which consists in studying configurations 
that are independent of one of the spatial coordinates, say $x_2$. 
Then, in order to deal with finite quantities that make sense, 
one must reinterpret ${\cal Q}$ as the supercharge per unit
length and also $M$ as a uniform longitudinal mass density. In
this case, the closed curve $\Gamma$ should be identified with a
discrete set of two points asymptotically located at both
infinities in the $x_1$-axis. Then, the equation (\ref{eq:central})
must be rewritten as
\begin{equation}
Z = {\cal W}(x_1\to\infty) - {\cal W}(x_1\to{-\infty}) \equiv
\Delta{\cal W} ~,
\label{eq:central2}
\end{equation}
and the supercharge algebra (\ref{eq:algebra}) should be understood as
the $N=2$, $d=2$ supersymmetry algebra which, in fact, admits the
appearance of central extensions. Now, the positive-definitness
of the supercharge algebra (\ref{eq:algebra}) implies
\begin{equation}
M \geq |\Delta{\cal W}| ~.
\label{eq:bound}
\end{equation}
This is nothing but the Bogomol'nyi bound of the coupled real 
scalar system introduced above. Indeed, although $\Delta{\cal W}$
is apparently different from the usual definition of the topological
charge $T$, it actually coincides with it since both depend only on
the topology \cite{wo}. It can be seen that $\Delta{\cal W}$ vanishes
in a topologically trivial state and has a positive value in a 
domain ribbon state. The appearance of the topological charge as
a central extension of the supercharge algebra seems to contradict
previous results obtained in Ref.\cite{hs} for kink states in
$d=2$ systems. However, we must point out that the configurations
we are considering fit into a dimensional reduction scheme,
and so the result (\ref{eq:central2}) can be seen as an expected
result \cite{ens,hs2}.

~

It is straightforward to see that the configurations that
saturate the bound (\ref{eq:bound}) are those preserving one half of
the supersymmetries. The explicit result appears after choosing
$\gamma^0\eta_{\pm}=\pm\eta_{\pm}$, which allows writing 
\begin{equation}
\{{\cal Q}[\eta_{\pm}],{\cal Q}[\eta_{\pm}\}\vert=
\half\int d^2x \left[ 
(\delta_{\eta_{\pm}}\psi)^{\dagger}(\delta_{\eta_{\pm}}\psi)+
(\delta_{\eta_{\pm}}\rho)^{\dagger}(\delta_{\eta_{\pm}}\rho)
\right] ~,
\label{eq:algebra2}
\end{equation}
and therefore the bound is saturated provided $\delta_{\eta_{+}}
(\psi,\rho) = 0$ or $\delta_{\eta_{-}} (\psi,\rho) = 0$, thus 
preserving the purely bosonic nature of the background configuration.
Here we recall that the only configuration that preserves all the
supersymmetries is the trivial vacuum configuration. Furthermore,
the equations that saturate the lower bound  are nothing but the
Bogomol'nyi equations of the system:
\begin{equation}
\frac{d\phi}{dx} \pm \frac{\partial{\cal W}}{\partial\phi} = 0 ~,
\label{eq:bog1}
\end{equation}
\begin{equation}
\frac{d\chi}{dx} \pm \frac{\partial{\cal W}}{\partial\chi} = 0 ~.
\label{eq:bog2}
\end{equation}
Let us consider the case when the supersymmetry corresponding to
the parameter $\eta_+$ is unbroken (thus, eqs.(\ref{eq:bog1})
and (\ref{eq:bog2}) are valid with the upper sign). The
supersymmetry generated by $\eta_-$ is broken in the BPS domain
ribbon background given by the solution of these equations.
The variations (\ref{eq:traf2}) and (\ref{eq:traf3}) for the
broken supersymmetries, as expected, give zero energy Grassmann 
variations of the domain ribbon solution; that is, they are zero 
modes of the Dirac equation in the background of the defect as can 
be easily verified. Quantization of these fermionic Nambu-Goldstone 
zero modes leads to the construction of
a (BPS) supermultiplet of degenerate bosonic and fermionic soliton 
states which transform according to a short representation of the
supersymmetry algebra. 

~

We have so far given a supersymmetry derived proof of the
existence of a Bogomol'nyi bound and self-duality equations
in the family of relativistic systems of coupled real scalar
fields first introduced in Ref.~\cite{bds95}.
In the next Section we will introduce specific systems, which 
comprise very interesting topological defects that may or may not 
present internal structure provided the superpotential ${\cal W}$ 
is conveniently choosen.

\section{Some Specific Systems}
\label{sec:systems}

Let us now consider explicit examples of systems of two real scalar
fields containing up to the sixth power in the scalar fields. To
illustrate the procedure, in the following we will consider three
different systems, the first two containing different powers in
each one of the fields. This is interesting because we will find
defects of different nature corresponding to each one of the two
fields. In the third system, both fields $\phi$ and $\chi$ enter
the game with equal footing, that is, the potential in this case
contains sixth order powers in both fields.

\subsection{BPS $\phi^6$ ribbons without internal kinks}
\label{subsec:system1}

As a first system to study in order to obtain a deeper insight
on the topological defects that result from the saturation of
the Bogomol'nyi bound, we will consider a superpotential of
the form
\begin{equation}
{\cal W}_{1}(\phi,\chi) = \half\l\phi^2\left(\half\phi^2
- a^2\right) + \half\mu\phi^2\chi^2 ~.
\label{eq:model1}
\end{equation}
The potential that results from it can be obtained after
(\ref{eq:pot}) to be
\begin{eqnarray}
U_{1}(\phi,\chi) & = & \half\l^2\phi^2(\phi^2 - a^2)^2 +
\lambda\mu\phi^2(\phi^2 - a^2)\chi^2 + \nonumber\\
& & \frac{1}{2}\mu^2\phi^2\chi^4 + \half\mu^2\phi^4\chi^2 ~.
\label{eq:pot1}
\end{eqnarray}
This potential has an explicit discrete symmetry $Z_2 \times Z_2$, and
degenerate vacuum states ($\phi^2=a^2$,~$\chi=0$) that breaks
the symmetry $Z_2$ corresponding to the $\phi$ field. It also has a flat
direction for the vacuum expectation value of $\chi$ when the scalar
field $\phi$ seats at the symmetric vacuum $\phi=0$. In particular,
the minimum ($\phi=0$,~$\chi=0$) preserves the whole
$Z_2 \times Z_2$ symmetry.

~

The set of first order differential (Bogomol'nyi) equations 
corresponding to bosonic configurations of this system is given by
\begin{equation}
\frac{d\phi}{dx} = \l\phi(\phi^2 - a^2) + \mu\phi\chi^2 ~,
\label{eq:boguno1}
\end{equation}
and
\begin{equation}
\frac{d\chi}{dx} = \mu\phi^2\chi ~.
\label{eq:bogdos1}
\end{equation}
The bosonic sector of this system was already investigated in \cite{bds95}. 
There it was found the following set of solutions: The first pair of 
solutions is $\chi=0$, and 
\begin{equation}
\phi^2(x) = \frac{1}{2}a^2[1-\tanh(\lambda a^2 x)] ~.
\label{eq:soluno1}
\end{equation}
It is easy to see that (\ref{eq:soluno1}) is a solution just by noting 
that, if one sets $\chi\to 0$ in the potential, one gets
\begin{equation}
U_{1}(\phi,0)=\frac{1}{2}\lambda^2\phi^2(\phi^2-a^2)^2 ~,
\label{eq:pot1bis}
\end{equation}
a sixth-order potential which is known to admit solutions of the form
(\ref{eq:soluno1}). 
The BPS domain ribbon is located at $x=0$ and its thickness
is given by $\delta \approx (\l a^2)^{-1}$. It is convenient to regard the 
domain ribbon as a slab of false vacuum of width $\delta$ with $\phi^2 =
a^2/2$ in the interior and $\phi^2 = 0$, $\phi^2 = a^2$ at
both sides of it.

~

A second pair of BPS solutions for the system above is
\begin{equation}
\phi^2(x) = \half a^2 [1 - \tanh(\mu a^2 x)] \;\;\;\;\; \mbox{and}
\;\;\;\;\; \chi^2(x) = \left(\frac{\l}{\mu} - 1\right)\phi^2(x) ~,
\label{eq:soldos1}
\end{equation}
and in this last case one must require that $\lambda/\mu>1$. Both 
solutions have the same energy per unit length, which is given by
\begin{equation}
E_1 = \frac{1}{4}|\lambda|a^4 ~,
\label{energy1}
\end{equation}
and this follows from the fact that both pairs belong to the same 
topological sector. Furthermore, both pairs of solutions are classically
or linearly stable, and this follows from the general result \cite{bsa96}
that ensures stability of solutions that solve the pair of first-order
differential equations, that is, of BPS solutions.

~

For the kink described by (\ref{eq:soluno1}), let us first point out that
this defect connects two regions which are very different from the
beginning: a symmetric region with a vanishing value of $\phi$ is connected 
to an asymptotic region where the discrete symmetry $Z_2$
related to the transformation $\phi \to -\phi$ is broken. In this sense,
this kind of solitons are asymmetric, in contrast to the $\phi^4$ system,
where the kink connects asymptotic regions both having asymmetric
vacua \cite{brs96b}. The issue here is that the $\phi^4$ model is usually
considered to simulate second-order phase transitions, where the kink defects
describe order-order interfaces that appear in this case. However, the
$\phi^6$ system is related to first-order transitions in which the two phases,
symmetric and asymmetric or disordered and ordered, may appear simultaneously.
Evidently, in this last case the kink defects describe order-disorder
interfaces. As we are going to show, interesting physical consequences
may appear in this richer situation where the potential can present up to
the sixth power in the fields. To see how this comes out, let us introduce
masses for the $\phi$ mesons living outside the kink:
$m^2_{\phi}(0,0)=\l^2a^4$ and $m^2_{\phi}(a^2,0)=4\lambda^2 a^4$,
depending upon the side. Furthermore, for $\phi=0$ the $\chi$ mesons
appear to be massless, and  for $\phi^2=a^2$ we obtain
\begin{equation}
U_{1}(\pm a,\chi) = \half\mu^2 a^4\chi^2 + \half\mu^2 a^2\chi^4 ~,
\label{eq:pota}
\end{equation}
and the $\chi$ mesons are such that $m^2_{\chi}(a^2,0)=\mu^2a^4$.
As a consequence, this system presents the interesting feature of
containing a topological defect separating the outside regions into two
distinct regions, one containing massless $\chi$ mesons and massive $\phi$
mesons with mass $\lambda^2a^4$, and the other
with massive $\chi$ and $\phi$ mesons, with masses $\mu^2a^4$ and
$4\lambda^2 a^4$, respectively. Because of this asymmetry, that makes $\chi$
mesons to be massless at the symmetric vacuun $\phi=0$, there is no way of
making the $\phi$ field to trap the $\chi$ field in its interior, to give
rise to a topological defect inside a topological defect in two space
dimensions. Since the $\chi$ field is massless
for $\phi=0$, there is no other energetic argument left to favor the
$\chi$ field to be inside the ribbon. For instance, the region inside the
ribbon is defined with $x=0$ and here we get $\phi^2=a^2/2$, which
changes the above potential to the form 
\begin{equation}
U_1(\pm a/\sqrt{2},\chi)=\frac{1}{4}\mu^2a^2\Biggl[
\chi^2-\frac{1}{2}\left(\frac{\l}{\mu}-
\frac{1}{2}\right)a^2\Biggr]^2+
\frac{1}{16}\mu^2a^6\left(\frac{\l}{\mu}-\frac{1}{4}\right)~,
\end{equation}
which presents spontaneous symmetry breaking for $\l/\mu>1/2$.
However, spontaneous symmetry breaking requires the presence of
massive $\chi$ mesons inside the ribbon, and these massive mesons
would instead decay into the massless mesons that live outside the
ribbon. We remark that the case $\l/\mu=1/2$ seems to be interesting
since this would also make the $\chi$ mesons to be massless inside the
domain ribbon, but here spontaneous symmetry breaking would unfortunately
not be present anymore.

~

Before ending this subsection, let us comment a little on stability by
following the standard way. From the above potential $U_1(\pm a/\sqrt{2},\chi)$
we can write the corresponding kink solutions, for $\l/\mu>1/2$,
\begin{equation}
\chi=\pm\sqrt{\frac{1}{2}\left(\frac{\l}{\mu}-\frac{1}{2}\right)}\,\,a
\tanh\Biggl[\frac{1}{2}\mu\sqrt{\frac{\l}{\mu}-
\frac{1}{2}\,}\,\,a^2\,\,y\,\Biggr]~.
\end{equation}
These solutions, or better the pairs of solutions given by $\phi^2=a^2/2$
and $\chi$ as above, do not solve the first-order equations and so do not
give any BPS solution. For this reason the proof of classical stability
already introduced in \cite{bsa96,brs96a} does not work for them. To
investigate stability we should consider
\begin{equation}
\phi(y,t)=\bar{\phi}+\sum_n\eta_n(y)\, \cos(w_nt)~,
\end{equation}
and
\begin{equation}
\chi(y,t)=\bar{\chi}+\sum_n\zeta_n(y)\, \cos(w_nt)~,
\end{equation}
where $\bar{\phi}$ and $\bar{\chi}$ constitute the pair of classical
solutions and $\eta(y)$ and $\zeta(y)$ are small fluctuations.
In this case we substitute the above configurations into the equations
of motion to get to the equation
\begin{equation}
S\,\,\left({\eta_n\atop\zeta_n}\right) = 
w_n^2\,\,\left({\eta_n\atop\zeta_n}\right) ~,
\end{equation}
valid for small fluctuations and for static classical field configurations.
Here $S$ is the Schr\"odinger-like operator
\begin{equation}
S=-\frac{d^2}{dy^2}+V(y)~,
\end{equation}
where $V(y)$ is a $2\times2$ matrix with elements $V_{ij}=
\partial^2V/\partial f_i\partial f_j$, where $f_1=\phi$ and $f_2=\chi$.
Evidently, $V(y)$ is to be calculated at the static classical configurations
$\bar{\phi}$ and $\bar{\chi}$ where the small fluctuations are being considered.

~

For the BPS pair of solutions we have $\chi=0$; this decouples the
fluctuations and allows introducing an explicit analytical investigation,
as already done in \cite{bds95}. For the above non-BPS pair of solutions,
however, neither $\bar{\phi}$ nor $\bar{\chi}$ vanishes, and so $V(y)$ does
not become diagonal anymore. To get an idea here let us just write the
non-diagonal elements of $V(y)$ in this case:
\begin{equation}
V_{12}=V_{21}=\mu^2a^4 f(\l/\mu)\tanh[g(y)]\{1+f^2(\l/\mu)\tanh^2[g(y)]\}~,
\end{equation}
with $f(\l/\mu)=\sqrt{\l/\mu-1/2\,}$ and $g(y)=(1/2)\mu f(\l/\mu) a^2 y$.
Experience on former investigations \cite{bnr97} says that the resulting
Schr\"odinger equation does not even map the exactly solvable modified
Posch-Teller problem \cite{mf}, and so a standard investigation concerning
classical stability can only be implemented numerically, but this is out of
the scope of the present work. Such an investigation should confirm 
instability of that pair of non-BPS solutions, as suggested by the energy 
considerations presented above.

\subsection{$\chi^4$ kinks inside a BPS $\phi^6$ ribbon}
\label{subsec:system2}

Let us now consider another $\phi^6$ system. We will show in this 
subsection that it is possible to circumvent the asymmetry outside 
the domain ribbon of the $\phi$ field that we have just found in the 
above system, in spite of the fact that the solution itself is asymmetric. 
Moreover, we will find that the resulting domain ribbon admits a 
non-trivial internal structure that can be shown to be a kink.
In order to see this, we consider a system described by a modified 
superpotential ${\cal W}_{2}$ given by
\begin{equation}
{\cal W}_{2}(\phi,\chi) = \half\l\phi^2\left(\half\phi^2 - a^2\right) + 
\half\mu\left(\phi^2 - \half a^2\right)\chi^2 ~.
\label{eq:model2}
\end{equation}
This superpotential gives the following scalar potential
\begin{eqnarray}
U_{2}(\phi,\chi) & = & \half\l^2\phi^2(\phi^2 - a^2)^2 +
\l\mu\phi^2(\phi^2 - a^2)\chi^2 + \nonumber\\
& & \half\mu^2\phi^2\chi^4 + \half\mu^2\left(\phi^2 - \half 
a^2\right)^2\chi^2 ~,
\end{eqnarray}
that present a quite different vacuum structure without 
flat directions. The degenerate vacua are given by the fully
symmetric state ($\phi^2=0$,~$\chi^2=0$), a couple of $Z_2$
invariant states ($\phi^2=a^2$,~$\chi^2=0$) and a set of states
($\phi^2=a^2/2$,~$\chi^2=(\l/2\mu)a^2$) that break the whole
$Z_2 \times Z_2$ symmetry.

~

For minimum energy, static field configurations obey the set of 
first-order equations
\begin{equation}
\frac{d\phi}{dx} = \lambda\phi(\phi^2 - a^2) + \mu\phi\chi^2 ~,
\label{eq:boguno2}
\end{equation}
and
\begin{equation}
\frac{d\chi}{dx} = \mu\left(\phi^2 - \half a^2\right)\chi ~,
\label{eq:bogdos2}
\end{equation}
which present soliton solutions. Indeed, if we set $\chi=0$, it is 
immediate that this system admits the BPS $\phi^6$ solution that we 
found earlier (\ref{eq:soluno1}), that is,
\begin{equation}
\phi^2(x) = \half a^2 [1 - \tanh(\lambda a^2 x)] ~.
\label{eq:soluno2}
\end{equation}
Another pair of solutions can be found using the trial orbit method
\cite{r79} to be
\begin{equation}
\phi^2(x) = \quart a^2 [1 - \tanh(\lambda a^2 x)] \;\;\;\;\; \mbox{and}
\;\;\;\;\; \chi^2(x) = \half \phi^2(x) ~,
\label{eq:soldos2}
\end{equation}
provided the coupling constants obey the relation $\mu=2\l$. The above two
pair of solutions are BPS solutions and, thus, are stable \cite{bsa96}.
As in the former asymmetric case, the $\phi^6$ domain ribbon connect regions
with different transformation properties under the symmetry subgroup $Z_2$.
However, if one computes the mass of the $\chi$ mesons at both
sides of the domain ribbon (\ref{eq:soluno2}), 
it is interesting to see that 
\begin{equation}
m^2_{\chi}(0,\chi) = m^2_{\chi}(a^2,\chi) = \quart \mu^2 a^4 ~,
\label{eq:masym}
\end{equation}
i.e., the $\chi$ field does not distinguish between the two 
outside regions of the BPS $\phi^6$ ribbon. This new scenario allows
building topological defects inside a topological defect. Indeed, it can
be seen that inside the domain ribbon we have $x=0$ and this makes
$\phi^2=a^2/2$. Consequently, the
$\chi$ mesons feel an effective potential given by
\begin{equation}
U_2(a^2/2,\chi) = \quart\mu^2 a^2\left(\chi^2 - \frac{\l}{2\mu}a^2\right)^2 ~.
\label{eq:potbis2}
\end{equation}
For $\lambda/\mu >0$ there is a spontaneous symmetry breaking 
potential for the $\chi$ field with vacua states located at $\chi^2=\l 
a^2/2\mu$. Moreover, the mass of the excitations is
\begin{equation}
m^2_{\chi}(a^2/2,\chi) = \l\mu a^4 ~,
\label{eq:masym2}
\end{equation}
and one sees that $m^2_{\chi}(out) > m^2_{\chi}(in)$ for
$\lambda/\mu< 1/4$, as follows from the above results. This restriction on
the parameters ensures that the $\phi^6$ ribbon traps the $\chi$ kink in $2+1$
dimensions. Indeed, it is energetically favorable for the $\chi$
bosons to remain inside the ribbon: the mass of the boson field $m(x)$
increases away from the center of the ribbon, resulting in a force $F 
\approx -\partial{m(x)}/\partial{x}$ that attracts $\chi$ bosons toward 
the ribbon, entrapping them. Here we remark that the $\phi$ defect is of 
the $\phi^6$ type, while the $\chi$ defect is of the $\chi^4$ type, and 
so they are of different nature.

~

In the present case there exists a parameter range $\l/\mu\in(0,1/4)$,
in which it becomes energetically favorable for a $\chi$ condensate to form
within the core of the domain ribbon. Indeed, the potential
(\ref{eq:potbis2}) is minimized by a field configuration for which
$\chi = \pm\chi_0$ where
\begin{equation}
\chi_0 = \left(\frac{\l}{2\mu}\right)^{1/2}a ~.
\label{eq:chio}
\end{equation}
We take for simplicity $a$ real and positive. It is straightforward to see
that $U_2(a^2/2,\chi_0) < U_2(a^2/2,0)$, and this shows that inside the
domain ribbon associated to the $\phi$ field it is possible that domains 
where $\chi = \pm\chi_0$ appear. It is clear that domains with $\chi_0$ and
$-\chi_0$ should necessarily be connected by topological defects.
The interior of the BPS domain ribbon is then a region where the
discrete symmetry $Z_2$ associated with the $\chi$ field is broken.
Thus, inside the domain ribbon scalar condensates will eventually form,
but they will be uncorrelated beyond some coherence length $\xi$ \cite{mor95}.
We therefore expect domains of $\chi = +\chi_0$ and $\chi = -\chi_0$
to form at different positions along the domain ribbon, with each domain
extending an average length given by $\xi$. Different domains must be
separated by a region where $\chi = 0$ that should be understood
as the location of the resulting $\chi$ kink. This is an example of a
topological kink inside the topological domain ribbon.
The explicit form of the solution is given by (\ref{eq:soluno2}) for
the host domain ribbon, whereas for the kink we just have
\begin{equation}
\chi(y)=\sqrt{\frac{\lambda}{2\mu}}\,a
\tanh\left(\frac{1}{2}\sqrt{\lambda\mu}\,a^2 \,y\right) ~.
\end{equation}
Evidently, the domain ribbon appears from the pair of BPS solutions
given by $\chi=0$ and $\phi$ as in Eq.~{$(\ref{eq:soluno2})$}, and is a ribbon,
a defect of dimension one in the planar system that we are considering.
For the kink, however, we see that it appears {\em inside} the domain ribbon,
which is located at $x=0$ and extends along the direction described by $y$ in 
the present case. The kink is a topological
defect of dimension zero and appears after setting $\phi^2=a^2/2$ and removing
the $x$ degree of freedom in the above system.

~

Investigations concerning classical stability may be introduced by just
following the steps already presented in the former subsection. Like there, no
analytical result can be obtained in the present case too. Here, however, the
modification introduced in the potential allows the appearence of a region
in parameter space obeying $0<\l/\mu<1/4$, in which the above pair of
solutions might be classically or linearly stable. In this region, as one knows 
two consecutive kinks must be separated by an antikink, so that
the initial kink-antikink separation distance should be of the order of the
coherence length $\xi$. To determine the explicit form of this coherence
length we see that the width of the above kink solution is given by
$1/\sqrt{\l\mu}a^2$, which essentially measures the distance between
consecutive domains of $\chi^2=\chi_0^2$, and so should be identified with
the coherence length in the present case, that is, $\xi \approx
1/\sqrt{\l\mu}a^2$.

\subsection{$\chi^6$ kinks inside a BPS $\phi^6$ ribbon}
\label{subsec:system3}

Let us now consider a third system of coupled real 
scalar fields that, in spite of displaying an asymmetry of the effective
potential outside the domain ribbon, allows the formation of a 
non-trivial internal structure inside the defect. As we shall see, this
system presents a structure that gives a kink of the $\chi^6$ type, that is,
of the same nature of the host domain ribbon. To see how this works explicitly,
let us consider the following superpotential
\begin{equation}
{\cal W}_{3}(\phi,\chi) = \half\l\phi^2\left(\half\phi^2
- a^2\right) - \half\mu\chi^2\left(\frac{1}{8}\chi^2 ~ - a^2\right) 
+ \half\mu\phi^2\chi^2 ~.
\label{eq:model1b}
\end{equation}
Here we see that both fields are very similar, although the symmetry is still
$Z_2 \times Z_2$. The scalar potential
generated by this superpotential is given by 
\begin{eqnarray}
U_{3}(\phi,\chi) & = & \half\l^2\phi^2(\phi^2 ~ - a^2)^2 +
\half\mu^2 \chi^2\left(\frac{\chi^2}{4} ~ -a^2\right)^2 + 
\frac{1}{4}\mu^2\phi^2\chi^4 + \nonumber\\
& & \half\mu^2\left( 1+ \frac{2\lambda}{\mu}\right)\phi^4\chi^2
+ \mu^2\left( 1 - \frac{\lambda}{\mu}\right)a^2\phi^2\chi^2 ~.
\label{eq:pot1b}
\end{eqnarray}
It is clear from the above expression that this system displays
spontaneous symmetry breaking for each one of the two field, separately.
To see this explicitly, we note that
\begin{equation}
U_{3}(\phi,0)=\frac{1}{2}\lambda^2\phi^2(\phi^2-a^2)^2~,
\end{equation}
and
\begin{equation}
U_{3}(0,\chi)=\frac{1}{2}\mu^2\chi^2\left(\frac{1}{4}\chi^2-a^2\right)^2~.
\end{equation}
Thus, a domain ribbon solution of the type previously studied exists for 
each one of the non-vanishing fields. In order to see which field the
system chooses to host the kink, we have to investigate the
energy of kinks generated by the corresponding $1+1$ dimensional
systems described by the above potentials.
For a more complete investigation concerning this point see 
Ref.~\cite{bba97}, where the high temperature effects on the class of 
systems of interest are presented\footnote{These thermal effects are
relevant to the standard cosmological scenario for the formation of
the host domain ribbons, since one knows that the cosmic evolution
occurs via expansion and cooling.}. For the system defined by $U_3(\phi,0)$ 
the energy of the corresponding $\phi^6$ kink is
\begin{equation}
E_{\phi-\mbox{\scriptsize{kink}}} = \frac{1}{4}|\lambda| a^4~.
\end{equation}
For the system defined by $U_3(0,\chi)$ we get
\begin{equation}
E_{\chi-\mbox{\scriptsize{kink}}} = |\mu| a^4~,
\end{equation}
and so we can write the result
\begin{equation}
E_{\phi-\mbox{\scriptsize{kink}}} =
\frac{|\lambda|}{4|\mu|}E_{\chi-\mbox{\scriptsize{kink}}}.
\end{equation}
Thus, if we choose to work with $|\lambda|\ge4|\mu|$ we can consider
the $\phi$ field as the field to generate the host domain ribbon.

~

In general, for static solutions the first order equations are given by
\begin{equation}
\frac{d\phi}{dx} = \l\phi(\phi^2 - a^2) + \mu\phi\chi^2 ~,
\label{eq:boguno1b}
\end{equation}
and
\begin{equation}
\frac{d\chi}{dx} =- \mu\chi\left(\frac{\chi^2}{4} -a^2\right) + 
\mu\phi^2\chi ~,
\label{eq:bogdos1b}
\end{equation}
and the domain ribbon appears after setting $\chi\to 0$. Let us examine the 
behavior of the $\chi$ field in the background of this defect.
We take the parameters $\lambda$ and $\mu$ real and positive with 
$\lambda = 4\mu$, for simplicity. In this case, the $\chi$ field can
generate a kink inside the BPS domain ribbon. In fact, $\phi^2=a^2/2$
in the core of the domain ribbon, and the corresponding effective 
potential results to be
\begin{equation}
U_3(a^2/2,\chi) = \mu^2a^6 
+\frac{1}{32}\mu^2\chi^2\left(\chi^2 - 2a^2\right)^2 ~.
\label{eq:ribb1b}
\end{equation}
Here a mass for the $\chi$ field can be introduced; it is given by
\begin{equation}
m^2(a^2/2, \chi) = \frac{\mu^2a^4}{4}
\label{eq:masb}
\end{equation}
Outside the domain wall, for $\phi \to 0$ the $\chi$ boson mass is given 
by $m_{\chi}^2(out) = \mu^2a^4$, and in the other outside region, where 
$\phi^2 = a^2$, the $\chi$ boson acquire a mass $m_{\chi}^2(out)=4\mu^2a^4$.
Then, for $\lambda = 4\mu$ we see that the $\phi^6$ ribbon traps the $\chi$ 
field and topological defects associated with the field $\chi$ will form
inside this domain ribbon. The situation here is different from the two
former cases, but the fact that $m_{\chi}^2(out) > m_{\chi}^2(in)$
ensures the presence of the $\chi$ particles inside the BPS ribbon.

~

The effective potential (\ref{eq:ribb1b}) is minimized by a field 
configuration for which $\chi = 0$ or $\chi = \pm\sqrt{2}a$, and now inside
the domain ribbon generated by the $\phi$ field it is possible that the $\chi$
field presents domains with $\chi = 0$ and $\chi^2 = 2a^2$. We recall that
this case is different from the former case, where the nested field was
governed by some $\chi^4$ potential. Here the $\chi$ field may present
symmetric and asymmetric domains, which should be connected by topological
defects. We expect domains with $\chi=\sqrt{2}a$,
$\chi=0$ and $\chi=-\sqrt{2}a$  to form at different positions,
uncorrelated beyond a given coherence length $\xi$. Correlated domains
should form, however, in conformity with the kink structure of the $\chi^6$
system, in which energy considerations favor the presence of kinks that
connect asymmetric vacua to the symmetric vacuum. To write the explicit
solutions we see from $U_3(\phi,0)$ that the host domain ribbon appears from
$\chi=0$ and
\begin{equation}
\phi^2(x) = \frac{1}{2}a^2[1-\tanh(\lambda a^2 x)] ~,
\label{eq:soluno3}
\end{equation}
and here we should set $\l\to 4\mu$, to get to the case we are considering
above. For the kink, we consider $U_3(a^2/2,\chi)$ and work on the transverse
direction to obtain
\begin{equation}
\chi^2(y) = a^2\left[1-\tanh\left(\half \mu a^2 y\right)\right] ~.
\label{eq:soluno4}
\end{equation}
These are the configurations for the host domain ribbon and the internal
kink in the present case, respectively. The initial separation distance 
between defects should be of the order $\xi$, the coherence length that is now given
by $\xi \approx 1/|\mu| a^2$. Like in the former case, here we can also have a region in
parameter space where stable kinks appear inside domain ribbons, and this is
another example where a BPS domain ribbon hosts topological kinks.

\subsection{Fermionic Behavior}
\label{subsec:fermions}

Let us investigate the presence of fermions in the background of the
BPS domain ribbon built from the $\phi$ field in the systems
introduced in the former subsections. To this end, we have to read
the effective mass of the fermions in the domain ribbon background from
the Yukawa couplings.

~

For the model of Sec.~4.1, the effective mass
of the fermions in the BPS domain ribbon background (\ref{eq:soluno1})
can be read off from the Yukawa couplings
\begin{equation}
{\cal L}^1_{Y} = - \frac{3}{2}\l\left(\phi^2(x) - 
\frac{a^2}{3}\right)\bar\psi\psi - \half\mu\phi^2(x)\bar\rho\rho ~.
\label{eq:yukawa}
\end{equation}
From this expression we see that it is energetically favorable for
the fermions $\psi$ to reside inside the domain ribbon, whereas this is
not the case for the other Majorana field. This result may perhaps lead
to a mechanism for finding charged domain ribbons, provided it
resists the complexification of the spinor $\psi$.

~

For the model of Sec.~4.2, the corresponding 
Yukawa couplings in the Lagrangian give
\begin{equation}
{\cal L}^2_{Y} = - \frac{3}{2}\l\left(\phi^2(x) - 
\frac{a^2}{3}\right)\bar\psi\psi - \half\mu\left(\phi^2(x) - 
\frac{a^2}{2}\right)\bar\rho\rho ~.
\label{eq:yukawa2}
\end{equation}
Thus, both fermions are trapped inside the BPS domain ribbon solution
corresponding to this system. Moreover, we would like to point out
that one of the Majorana fermions $\rho$ is massless in the core of
the defect and sees an isotropic exterior region. The possibility
of trapping massless fermions inside the domain ribbon may open an
interesting scenario: To find stable domain rings, the $2+1$ dimensional
analogue of domain bubbles in $3+1$ dimensions,  with the tension of the
domain defect being equilibrated by the quantum mechanical degeneracy
pressure exerted by the fermions that inhabit the ribbon.

~

The same analysis can be carried out in the system presented in
Sec.~4.3. In that case, the Yukawa couplings are given by
\begin{equation}
{\cal L}^3_{Y} = - \frac{3}{2}\l\left(\phi^2(x) - 
\frac{a^2}{3}\right)\bar\psi\psi - \half\mu\left(\phi^2(x) +
a^2\right)\bar\rho\rho ~.
\label{eq:yukawa3}
\end{equation}
From this expression we see that it is energetically favorable for
the fermions $\psi$ to reside inside the domain ribbon, whereas this is
not the case for the other Majorana field. Indeed, the fermion $\rho$
feels an effective potential that favor one of the exterior regions
of the domain ribbon. This result may perhaps lead to the formation
of Fermi disks, the $2+1$ dimensional analogue of Fermi balls in $3+1$
dimensions, which represent a bag of false vacuum populated by a Fermi gas
that stabilizes the soliton against collapse.

~

The above reasoning are mostly speculative and a more careful
investigation should be carried out, but this is out of the scope of
the present work. However, we recall that we have just studied the Yukawa
couplings in the background of the domain ribbon, for $\chi=0$. This
investigation is almost the same investigation one has to deal with in
systems with a single field --see, for instance, Ref.~{\cite{mba96}}--
for explicit calculations concerning Fermi balls in $3+1$ dimensions. 

\section{Comments and Conclusions}
\label{sec:comments}

In this work we have considered the existence of BPS topological defects
with internal structure in planar systems of two coupled real scalar
superfields. To this end, we have endowed the kind of systems of our
interest with an extended supersymmetry and looked into the sector of
extended solutions belonging to BPS representations. These
configurations are nothing but supersymmetric domain ribbons that
solve the set of Bogomol'nyi equations of the model. The system is
entirely defined in terms of just one function of the fields, the
superpotential ${\cal W}$. We have studied three different choices
for ${\cal W}$, some of them leading to the appearance of several seemingly 
interesting kinds of topological deffects: BPS domain ribbons
endowed with an internal structure given by a kink.

~

The first system is presented to illustrate the simplest case where
excitations of the bosonic field in the background of
a $\phi^6$ ribbon does not favor the formation of internal structure,
although it is favorable for the fermion $\psi$ to reside inside this
domain ribbon. This result may perhaps lead to a mechanism for finding 
charged domain ribbons, provided it resists the complexification of the 
spinor $\psi$. The presence of the domain ribbon is
shown to break one half of the supersymmetries, that is, it is a BPS
state. The superpartners of this configuration can be built just
by acting on it with the broken supercharges to obtain the fermionic
zero modes that shall be subsequently quantized.

~

The superpotential that leads to the second system is suggested by
symmetry considerations on the effective potential produced by the
background BPS ribbon. We have shown that the modified superpotential
allows the entrappment of bosons as well as the formation of $\chi$
kinks within the core of the domain ribbon. Here, however, the defects
are of different nature: The domain ribbon comes from the $\phi$
field (with a sixth-order potential) and connects the symmetric minimum 
$(\phi=0)$ to an asymmetric one $(\phi^2=a^2)$, while the kink is generated 
by the $\chi$ field (with a fourth-order effective potential) and
connects the two asymmetric minima $\chi^2=(\l/2\mu)\,a^2$. It is interesting
to recall that despite the asymmetric behavior of the field $\phi$ of the
ribbon at both outside regions, this asymmetry is not seen by the $\chi$ field
in this system. This new scenario allows building topological defects inside 
a topological defect. For a certain parameter range it becomes 
energetically favorable for $\chi$ condensates to form within the core of 
the domain ribbon. Different domains must be separated by a region where $\chi 
= 0$ that is naturally understood as the location of the resulting 
internal $\chi$ kink.

~

The third system is defined by another potential, in which both the $\phi$
and $\chi$ fields present quite a similar behavior. In this case we have
shown that the system admits defects of the same nature since now we can
find $\chi^6$ kinks inside a BPS $\phi^6$ domain ribbon, in spite of the
explicit asymmetry of the effective potential outside the domain ribbon.

~

The host domain ribbon solutions preserve one half of the supersymmetries,
thus being BPS states. They are known to be classically stable 
\cite{bsa96,brs96a}.
When we consider the inclusion of kinks inside these supersymmetric
domain ribbons, unfortunately, we cannot perform a complete analysis of 
their stability. 
Nevertheless, by means of simple energetic arguments, we are able to show
that there is a parameter range (in the second and third models) in which 
these solutions might be classically or linearly stable.

~

We have shown that the domain ribbon solution of the second system traps
both fermions and one of them become massless in the core of the
defect and sees an isotropic exterior region. This feature could lead to
the formation of stable domain rings of finite radius. On the other hand,
the third system has an effective potential for one of the Majorana fields
that favor one of the exterior regions of the domain ribbon. This result may 
perhaps lead to the formation of Fermi disks in $2+1$ dimensions. These
possibilities, as well as the finite temperature effects on the class of 
systems of our interest, deserve further investigations. We hope to
report on these issues elsewhere.

\section*{Acknowledgements}

The work of JDE and MLT was partially supported by Consejo Nacional de
Investigaciones Cient\'{\i}ficas y T\'ecnicas, CONICET. DB and FAB 
would like to thank Conselho Nacional de Desenvolvimento 
Cient\'{\i}fico e Tecnol\'ogico, CNPq, and Coordena\c{c}\~ao de Apoio
ao Pessoal do Ensino Superior, CAPES, for partial support and a
fellowship, respectively.


\begin{thebibliography}{99}
\bibitem{wit85} E.~Witten, Nucl. Phys. {\bf B249}, 557 (1985).
\bibitem{lsh85} G.~Lazarides and Q.~Shafi, Phys. Lett. B {\bf 159}, 261 
(1985).
\bibitem{mac88} R.~MacKenzie, Nucl. Phys. {\bf B303}, 149 (1988).
\bibitem{mor94} J.R.~Morris, Phys. Rev. D {\bf 49}, 1105 (1994).
\bibitem{mor95} J.R.~Morris, Phys. Rev. D {\bf 51}, 697 (1995).
\bibitem{bds95} D.~Bazeia, M.J.~dos Santos and R.F.~Ribeiro, Phys. Lett. 
A {\bf 208}, 84 (1995).
\bibitem{bsa96} D.~Bazeia and M.M.~Santos, Phys. Lett. A
{\bf 217}, 28 (1996). 
\bibitem{brs96a} D.~Bazeia, R.F.~Ribeiro and M.M.~Santos, Phys. Rev. E
{\bf 54}, 2943 (1996).
\bibitem{brs96b} D.~Bazeia, R.F.~Ribeiro and M.M.~Santos, Phys. Rev. D
{\bf 54}, 1852 (1996).
\bibitem{bmo97}D.~Bazeia and J.R.~Morris, {\it Domain walls and
nontopological solitons in systems of coupled fields.} Unpublished report.
\bibitem{comment} We remind that the supersymmetry algebra in $(2+1)$ 
spacetime dimensions allows the introduction of a real representation 
containing on-shell one bosonic and one fermionic degrees of freedom.
\bibitem{ggr} S.J.~Gates Jr., M.T.~Grisaru, M.~Ro\v{c}ek and W.~Siegel,
{\em One Thousand and One Lessons in Supersymmetry}, Frontiers in
Physics Vol.58, The Benjamin/Cummings Publishing Co., 1983.
World Scientific.
\bibitem{wo} E.~Witten and D.~Olive, Phys. Lett. B {\bf 78}, 97 (1978).
\bibitem{hs} Z.~Hlousek and D.~Spector, Nucl. Phys. {\bf B370}, 43
(1992).
\bibitem{ens} J.D.~Edelstein, C.~N\'u\~nez and F.A.~Schaposnik,
Phys. Lett. B {\bf 329}, 39 (1994).
\bibitem{hs2} Z.~Hlousek and D.~Spector, Nucl. Phys. {\bf B442},
413 (1995).
\bibitem{bnr97}D.~Bazeia, J.R.S.~Nascimento, R.F.~Ribeiro and D.~Toledo,
J. Phys. A {\bf30}, 8157 (1997).
\bibitem{mf} P.M.~Morse and H.~Feshbach, {\em Methods of Mathematical
Physics}, Mc Graw-Hill, New York, 1953.
\bibitem{r79}R.~Rajaraman, Phys. Rev. Lett. {\bf 42}, 200 (1979).
\bibitem{bba97} F.A.~Brito and D.~Bazeia, Phys. Rev. D {\bf 56}, 7869 (1997).
\bibitem{mba96}J.R.~Morris and D.~Bazeia, Phys. Rev. D {\bf 54}, 5217 (1996).
\end{thebibliography}
\end{document}